\documentclass[12pt,preprint]{aastex}

\newcommand{\FeH}{\mbox{[Fe/H]}\,}
\newcommand{\RRab}{\mbox{$RR{\mbox{\scriptsize ab}}\,$}}
\newcommand{\RRc}{\mbox{$RR{\mbox{\scriptsize c}}\,$}}
\newcommand{\RRd}{\mbox{$RR{\mbox{\scriptsize d}}\,$}} 

\shorttitle{The distance to the LMC cluster Reticulum}
\shortauthors{Dall'Ora et al.}

\begin{document}


\title{The distance to the LMC cluster Reticulum from the 
$K$-band Period-Luminosity-Metallicity relation of RR Lyrae 
stars\footnote{Based on observations collected at the European Southern 
Observatory, La Silla, Chile on INAF-Osservatorio Astronomico di Roma 
guaranteed time.}}


\author{M. Dall'Ora\altaffilmark{1}, J. Storm\altaffilmark{2}, 
G. Bono\altaffilmark{1}, V. Ripepi\altaffilmark{3}, M. Monelli\altaffilmark{1},
V. Testa\altaffilmark{1}, G. Andreuzzi\altaffilmark{1,4}, 
R. Buonanno\altaffilmark{1}, F. Caputo\altaffilmark{1}, 
V. Castellani\altaffilmark{1}, C. E. Corsi\altaffilmark{1}, 
G. Marconi\altaffilmark{5}, M. Marconi\altaffilmark{3}, 
L. Pulone\altaffilmark{1}, P. B. Stetson\altaffilmark{6}}


\altaffiltext{1}{INAF-Osservatorio Astronomico di Roma, via Frascati 33, 
00040 Monte Porzio Catone, Italy; fname@mporzio.astro.it}

\altaffiltext{2}{Astrophysikalisches Institut Potsdam, An der Sternwarte 16,
14482 Potsdam, Germany; jstorm@aip.de} 

\altaffiltext{3}{INAF-Osservatorio Astronomico di Capodimonte, via 
Moiariello 16, 80131 Napoli, Italy; ripepi/marcella@na.astro.it}

\altaffiltext{4}{Telescopio Nazionale Galileo, Istituto Nazionale di 
Astrofisica, P.O. Box 565, E-38700 Santa Cruz de La Palma, Spain; 
andreuzzi@tng.iac.es}

\altaffiltext{5}{European Southern Observatory, 3107 Alonso de Cordova, 
Santiago, Chile; gmarconi@eso.org}

\altaffiltext{6}{Dominion Astrophysical Observatory, Herzberg Institute of 
Astrophysics, National Research Council, 5071 West Saanich Road, Victoria, 
British Columbia V9E 2E7, Canada; Peter.Stetson@nrc-cnrc.gc.ca}


\begin{abstract}
We present new and accurate Near-Infrared $J$ and $K_s$-band data of the 
Large Magellanic Cloud cluster Reticulum. Data were collected with SOFI
available at NTT  and covering an area of approximately $5\times5$ arcmin 
around the center of the cluster. Current data allowed us to derive 
accurate mean $K$-band magnitudes for 21 fundamental (\RRab) and 9 
first overtone (\RRc) RR Lyrae stars. On the basis of the semi-empirical 
$K$-band Period-Luminosity-Metallicity ($PLZ_K$) relation we have recently 
derived, we find that the absolute distance to this cluster is 
$18.52\pm0.005 (random) \pm0.117 (systematic)$. Note that the current 
error budget is dominated by systematic uncertainty affecting the 
absolute zero-point calibration and the metallicity scale.
\end{abstract}


\section{Introduction}

Absolute distances of RR Lyrae stars play a fundamental role in stellar 
astrophysics, and the reasons are manifold. They are widely adopted to
establish the absolute age of globular clusters (GCs) and to properly compare 
predicted and empirical Color-Magnitude diagrams. Moreover, they are 
robust stellar tracers of low-mass, old populations and are ubiquitous 
across the Galactic spheroid as well as in Local Group galaxies.   
During the last few years a paramount theoretical and observational 
effort has been devoted to the RR Lyrae distance scale to improve not 
only the intrinsic accuracy but also to constrain the occurrence of 
deceptive systematic errors (Caputo et al. 2000; Walker 2000; Bono 2003; 
Cacciari \& Clementini 2003).   

In a series of previous papers (Bono et al. 2001,2002,2003, hereinafter 
B01, B02, and B03) we have drawn the attention to the $K$-band 
Period-Luminosity-Metallicity ($PLZ_K$) relation of RR Lyrae stars. 
This relation was originally discovered by Longmore et al. (1990, 
hereinafter L90) and presents several undisputable advantages when 
compared with other 
methods available in the literature. In particular, the $PLZ_K$ relation 
is only marginally affected by evolutionary effects as well as by a spread 
in stellar mass inside the instability strip. Moreover, it is only weakly  
affected by uncertainties in the reddening correction and shows a linear 
trend when moving from metal-poor to metal-rich objects. Current theoretical
predictions appear in very good agreement with empirical observations, and 
indeed the pulsation parallax to RR Lyr itself estimated by B02, is in very 
good agreement with the trigonometric parallax measured by 
Benedict et al. (2002) with the Fine Guide Sensor on board the Hubble Space 
Telescope (HST).   

Although the theoretical framework concerning the $PLZ_K$ relation has 
been thoroughly discussed, only a few investigations have been devoted 
to near-infrared (NIR) observations of cluster and field RR Lyrae stars. 
In fact, with the exception of the $K$-band data collected by L90 for cluster 
RR Lyrae and Baade-Wesselink RR Lyrae (see B03 for a detailed list) 
only one investigation has been devoted to RR Lyrae stars in the Galactic 
Bulge (Carney et al. 1995). The observational scenario has been recently 
complemented 
with new $K$-band data collected by Butler (2003), using the adaptive optics 
camera available on the 3.5m Calar Alto telescope, for a small sample of 
RR Lyrae stars located in the very center of the GC M3. The above 
discussion shows quite clearly that we lack new and accurate NIR data
to supply a sound empirical estimate of both the slope and the zero-point 
of the $PLZ_K$ relation but also to improve current empirical constraints 
on the metallicity dependence. To fill this gap we undertook an observational
project aimed at providing a comprehensive observational investigation of 
cluster and field RR Lyrae stars in the NIR $J$ and $K_s$-bands.  

In this investigation we present $K$-band data for RR Lyrae stars in the 
LMC cluster Reticulum. We selected this cluster for the following reasons:
it presents a sizable sample of well-studied RR Lyrae stars (32, Walker 1992), 
its central density is very low ($\log \rho_0\approx 0\; M_\odot pc^{-3}$, 
Peterson \& Kunkel 1977), and therefore the NIR photometry is not seriously 
affected by crowding, and the metal content 
of this cluster has been estimated using medium resolution spectra 
($[Fe/H]\approx=-1.71\pm0.1$, Suntzeff et al. 1992). In section 2, we 
present the observations and the fit to the $K_s$-band light curves, while 
in section 3 we discuss the new distance determination together with 
intrinsic and systematic uncertainties affecting the current estimate.

\section{Observations}

Near-Infrared $J$ and $K_s$ images of Reticulum were collected in three
different runs from December 1999 to February 2002 with SOFI available 
at NTT ESO La Silla. The seeing conditions during these observing runs 
were good and range from 0.5 to 1.3 arcsec. We collected 46 $J$ and 171 
$K_s$-band frames ($5\times5$ arcmin, pixel size=0.292 arcsec) centered 
on the cluster and 
the individual exposure times range for the $J$-band from 10 to 30 s 
and for the $K_s$-band from 10 to 120 s. 
We detected 30 out the 32 RR Lyrae stars present in this cluster, 
and the number of phase points for each object is approximately 12. 
Raw frames were 
pre-reduced using standard IRAF procedures. To improve the photometric 
accuracy along the light curves we  performed several tests using 
different techniques to stack the images, different criteria to select 
PSF stars across the individual frames as well as different strategies 
to perform the photometry over the entire data set. We found that the 
best intrinsic accuracy can be achieved if we do not stack the images. 
We selected a reasonable number of  PSF stars ($\approx$25) for 
each frame and adopted a variable PSF. Moreover, the NIR ($J$, $K_s$) 
frames have been simultaneously reduced with DAOPHOT/ALLFRAME together 
with a few V-band images to improve the accuracy of individual measurements 
and the limiting magnitudes. Optical (U,B,V,I) data have been collected 
during the same observing runs, with SUSI2 available at the same telescope. 
A detailed description of observing and data reduction strategies will be 
discussed in a companion paper.  
The absolute photometric calibration of $J$ and $K_s$ ESO magnitudes
into the LCO system was performed using the standard star 9109 (HST S055-D)
measured by Persson et al. (1998). We did not apply any correction to 
transform the $K_s$ into the $K$-band, due to the marginal difference 
(Lidman et al. 2002). 
This star was measured 10 times at air masses that bracket the 
observations on Reticulum and the accuracy of the calibration is of the 
order of 0.002 mag in the $K$-band.

Figure 1, shows the Color-Magnitude diagrams in the J, V-J (top) and 
$K$, $V-K$ (middle), and $V$, $B-V$ (bottom) bands together with the 
intrinsic photometric accuracy on the mean magnitudes provided by 
ALLFRAME. Data plotted 
in this diagram show quite clearly that the intrinsic accuracy for 
typical RR Lyrae luminosities reaches 0.01 at $J\approx 18.5$ and 
0.015 at $K\approx18$. We detected approximately 850 and 550 stars 
in $J$ and $K$-band, while the limiting magnitudes are $\approx20.5$ and 
$\approx20$, respectively. 
Note that the sample of hot stars located at $K\approx19.75$ and
$1.0 \le V-K\le 1.6$ are Blue Straggler stars as supported by the
cross-identification with the optical CMD (see open circles). 
This suggests that current photometry almost approaches the 
Turn-Off region.   

To further improve the intrinsic accuracy of the RR Lyrae mean magnitudes 
we decided to perform a fit of the individual phase points measured  
by ALLFRAME with a template curve. This approach allows us a better 
propagation of individual errors on the final mean magnitude and avoids 
the spurious fluctuations introduced by the binning of phase points.  
Jones et al. (1996) have developed a template fitting 
technique for determining mean magnitudes in the $K$-band for 
RR Lyrae stars from few phase observations. They have determined five 
different template shapes depending on the type of variability 
(\RRab or \RRc type) and on the $B$-amplitude of the variable and 
described each of these templates with a fourier series. We have 
adopted their method and for each star we computed the corresponding 
fourier template based on the ephemerides and the $B$-amplitudes 
from Ripepi et al. (2003, in preparation). The candidate \RRd stars 
have been treated as \RRc variables, due to their small amplitudes 
and relatively short periods. 

For each observed phase point we can compute the best estimate of the 
mean magnitude over phase of the variable by subtracting the template 
magnitude from the observed magnitude, 
$\langle m(\phi) \rangle = K_{obs}(\phi) - K_{template}(\phi)$. 
We computed the weighted intensity averaged magnitude over all the 
observed phase points for each variable star. The resulting 
mean magnitudes and their associated intrinsic errors are listed in Table 1. 
Figure 2 shows the mean  $\langle K \rangle$ magnitude as a function 
of $\log P$. Note that the period of \RRc variables (open circles) in 
Figure 2 have been fundamentalized, i.e. $\log P_F = 0.127 + \log P_{FO}$. 
It can be seen that there is an excellent correlation. Linear regression gives:

\begin{equation}
<K> = -2.16 (\pm 0.09) \log P + 17.352 (\pm0.025)
\end{equation}

with a standard deviation of only 0.03~mag.  This is a remarkable result 
for a standard candle.

\section{Discussion and final remarks}

Based on the mean $K$-magnitudes from the previous section we now
proceed to determine the distance to the Reticulum cluster. 
To further improve the theoretical calibration of the $PLZ_K$ relation,
B03 devised a new pulsation approach that relies on mean $K$-band magnitudes
and $V-K$ colors. In particular, they derived new $PLZ_K$ relations (see their
relations 3 and 4) and Period-Luminosity-Color-Metallicity, $PLCZ_{(V,K)}$,
relations (see their relations 7 and 8) that include the luminosity term.
Note that these relations when compared with theoretical predictions by B01
present three main advantages: {\em i)} they rely on a larger set of pulsation
models;  {\em ii)} independent theoretical relations have been derived
for fundamental and first overtone pulsators; {\em iii)} the new $PLZ_K$ and
$PLCZ_{(V,K)}$ relations were derived by adopting predicted mass values for
each assumed chemical composition, while B01 adopted an ensemble average.
To root the calibration of the zero-point as well as of the coefficient of
the metallicity term of the new $PLZ_K$ relation on empirical data, B03
adopted a sample of field RR Lyrae stars for which good $V$ and $K$-band
light curves, accurate reddening corrections, and metal abundances were
available in the literature. For fundamental mode pulsators they found:  

\begin{equation}
M_K = -0.770 (\pm 0.044) - 2.101 \log P + 0.231 (\pm 0.012) \FeH  
\end{equation}

Suntzeff et al. (1992) have determined the metallicity of the Reticulum 
cluster to be $\FeH=-1.71\pm0.1$ based on spectroscopy of the Calcium II 
triplet at around 860~nm for 9 individual stars in the cluster. 
Entering this value in equation 2 we can
determine the distance modulus, $(m-M)_K=\langle K \rangle - M_K$,
to each star. Forming the weighted average of these estimates we find
for the \RRab stars a value of $(m-M)_K=18.531\pm0.006$, while for the 
\RRc stars, using the fundamentalized periods, we find 
$(m-M)_K=18.539\pm0.009$ and for the complete sample we
find $(m-M)_K=18.534\pm0.005$.

The reddening towards Reticulum has been determined by Walker (1992) 
who finds a very low value of $E(B-V)=0.03\pm0.02$.
We combine this value with the reddening law from Cardelli, Mathis, 
\& Clayton (1989) of $A_K = 0.114 \times 3.1 \times E(B-V)$ which 
gives a minuscule
absorption in the $K$-band of $A_K = 0.011$~mag. Correcting the modulus
for the reddening we find a best estimate of $(m-M)_0=18.523\pm0.005$,
where the error estimate only consider intrinsic random errors.
The referee suggested to perform a detailed check of the systematic
uncertainties affecting distance estimates based on the semi-empirical
relation derived by B03. To accomplish this goal we selected the Galactic 
Globular Clusters (GGCs) for which are available both optical and NIR 
mean magnitudes. To avoid
subtle uncertainties in the reddening corrections we only selected GGCs
with reddening corrections smaller than 0.2. We ended up with four GGCs
whose metallicity ranges from $[Fe/H]=-1.27$ (M~5) to $[Fe/H]=-2.26$
(M~15). To estimate the difference in the distance moduli we adopted
distances based on the Baade-Wesselink (BW) calibration provided by
Fernley (1994) as well as on the First Overtone Blue Edge (FOBE)
provided by Caputo et al. (2000). Data listed in the last two columns
of Table 2 show that the difference ranges from $\Delta \mu=0.32$  
for the less metal-poor cluster (M~15) to $\Delta \mu=0.10$ for  
the more metal-rich cluster (M~5). 

This finding further strengthens the
result obtained by B03, i.e. the difference between distances based
on $PLZ_K$ and on the BW method depend on the mean metallicity and
it decreases in the metal-rich regime (see figures 6 and 8 in B03).
Unfortunately, we still lack accurate mean K-band magnitudes for
RR Lyrae stars in a larger sample of metal-rich GGCs. Moreover, 
current uncertainties affecting distance moduli based on mean V-band 
magnitudes are systematically larger than 0.1 mag. This hampers any 
precise quantitative conclusion on the systematic uncertainty affecting 
the $PLZ_K$ distance scale.

It is noteworthy, that cluster distances presented here appear to be
systematically larger than distances based on the FOBE method and on
the $PLZ_K$ calibration based on theoretical models provided by B01.
The semi-empirical calibration by B03 presents a steeper dependence
on the metallicity (0.231 vs 0.167) and the reasons for the difference
between the empirical and theoretical scales are not clear yet (see
also Carney et al. 1992).
Finally, we note that the current distance estimate to $\omega$ Cen
is also 0.1 mag larger than the distance modulus derived by Thompson
et al. (2001) on the basis of the eclipsing binary OGLE17. Although,
the analysis of current uncertainties affecting the RR Lyrae distance
is far from being definitive we decided to adopt a systematic 
uncertainty of 0.1 in the calibration of the zero-point.  

The uncertainty on the reddening correction is quite negligible, but 
we have to account for the uncertainty in the metallicity and in the 
metallicity scale. Equation 2 was derived using the metallicity 
estimates provided by Fernley et al. (1998) and by Zinn (1985) 
for field and cluster RR Lyrae stars. The metallicities of field 
RR Lyrae rely on a calibration of the $\Delta S$ method provided by 
Fernley \& Barnes (1997), i.e. $[Fe/H]=-0.13 - 0.195 \Delta S$, that is 
quite similar to the calibration provided by Clementini et al. (1995),
$[Fe/H]=-0.08 - 0.195 \Delta S$. These calibrations supply metal 
abundances quite similar to the Zinn \& West (1984) metallicity 
scale. On the other hand, Gratton (1999) provided a new calibration, 
$[Fe/H]=-0.03 - 0.176 \Delta S$, that relies on the Carretta \& 
Gratton (1997) metallicity scale. By accounting for these empirical 
uncertainties on the zero-point, on the slope as well as on the 
systematic uncertainties affecting the metallicity scale (Rutledge, 
Hesser, \& Stetson 1997; Kraft, \& Ivans 2003) we estimate the 
uncertainty on the metallicity to be of the order of 0.25 dex. 
This means that the uncertainty on the mean metallicity in equation 2
as well as in the cluster metallicity introduces an uncertainty of the
order of 0.06 mag in the current distance determination. Therefore, we
end up with a distance to Reticulum of $18.52\pm0.005 (random) \pm0.117
(systematic)$. 
   
Reticulum is located 11 degrees away from the center of LMC, and therefore
we can only supply weak constraints on the absolute distance to LMC itself.
According to our optical photometry and by assuming a mean reddening of
$E(B-V)=0.03\pm0.02$ (Walker 1992) the mean visual magnitude of RR Lyrae
stars in Reticulum is: $V_0=18.96\pm0.05$ mag. Note that this estimate is 
in very good agreement with the estimate provided by Walker (1992), i.e.
$V_0=18.98\pm0.04$. Recent estimates of the mean visual magnitude of RR Lyrae
stars in the LMC bar range from $V_0=19.05\pm0.06$ (108 stars) by Clementini
et al. (2003) to $V_0=18.90\pm0.02$ (7110 stars) derived by the OGLE team
(Soszynski et al. 2003) and to $V_0=18.99\pm0.02 (random)\pm 0.16 
(systematic)$ (80 first overtones) derived by the MACHO team 
(Alcock et al. 2004). The key differences in these estimates are:
statistics, absolute zero-point calibration, and reddening correction.
It goes without saying that according to these estimates the position of
Reticulum ranges from $\approx 3$ kpc in front of LMC to $\approx 2$ kpc
beyond LMC. New and homogeneous NIR data for a sizable sample of RR Lyrae
in the LMC bar are mandatory to settle the uncertainty on the relative
position of this cluster.

The empirical evidence brought out by current data suggests that the $PLZ_K$ 
relation is characterized by a global accuracy better than 0.12 mag. 
Note that the current discrepancy between HB evolutionary models constructed 
by adopting different assumptions concerning the input physics is of the 
order of 0.1-0.15 mag (Cassisi et al. 1999; VandenBerg et al. 2000). 
However, the most recent systematic survey of $K$-band data for cluster 
RR Lyrae stars dates back to L90. Therefore new and homogeneous NIR 
data for GGCs with sizable samples of RR Lyrae stars are strongly 
required to shed new lights on this {\em vexata questio}. At the same 
time, a new empirical calibration of the $PLZ_K$ relation based on  
RR Lyrae stars for which heavy-element abundances has been measured 
on high-resolution spectra is also necessary. 
    
The intrinsic accuracy of the $PLZ_K$ relation seems also very promising in 
providing a new distance scale for GCs that can allow us to improve their 
absolute age estimate. It is noteworthy that new cluster data will allow us 
to derive an independent $PLZ_K$ relation for first overtone RR Lyrae stars,
and in turn to further improve the intrinsic accuracy of this method. 
Finally, due to the marginal dependency on reddening corrections it could 
also be soundly adopted to derive the 3D-structure not only of the 
Magellanic Clouds, but also of the Galactic bulge.


\acknowledgments
This work was partially supported by MIUR/Cofin~2001, under the project 
"Galactic Spheroid", and MIUR/Cofin~2002, under the project "Stellar 
Populations in Local Group Galaxies".

\clearpage


\begin{figure}
\plotone{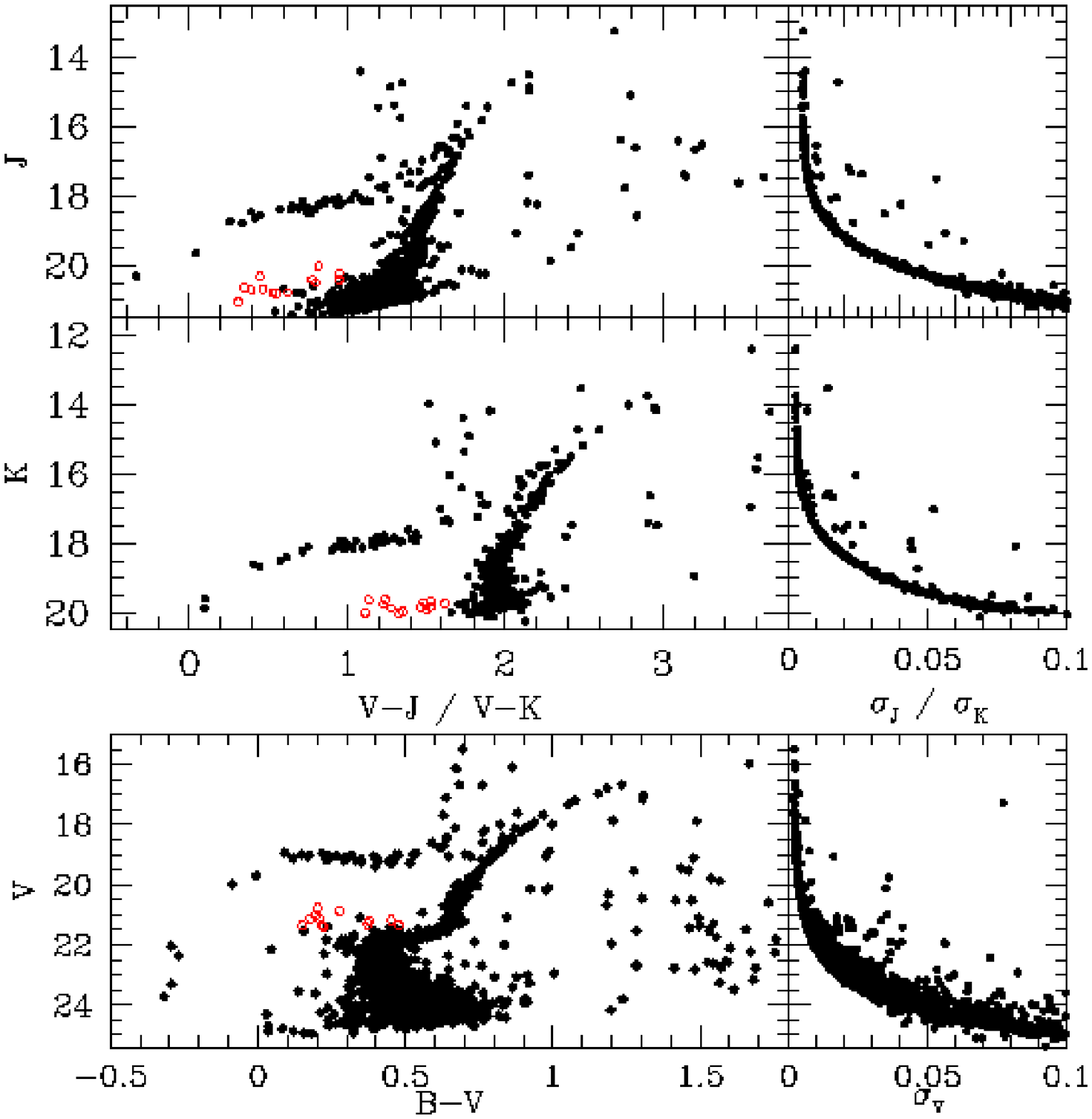}
\caption{Reticulum Color-Magnitude Diagrams in the $J$, $V-J$ (top), 
$K$, $V-K$ (middle), and $V$,$B-V$ (bottom) bands. The boxes on the right 
show the intrinsic photometric errors given by ALLFRAME. The NIR photometry 
is quite accurate, and indeed for $J\approx 18.5$ and $K_s\approx18$,
i.e. the typical  magnitudes of RR Lyrae stars, the intrinsic accuracy
is systematically better than 0.02. The open circles mark the position
of Blue Stragglers detected in the $K$, $V-K$ CMD and then cross-identified
in the other CMDs. By adopting the following selection criteria:
$|sharpness| \le 1.1$, and $\chi\le 0.5$ we detected approximately 850 
and 550 stars in the $J$ and in the $K_s$-band, respectively.\label{fig1}}
\end{figure}

\begin{figure}
\plotone{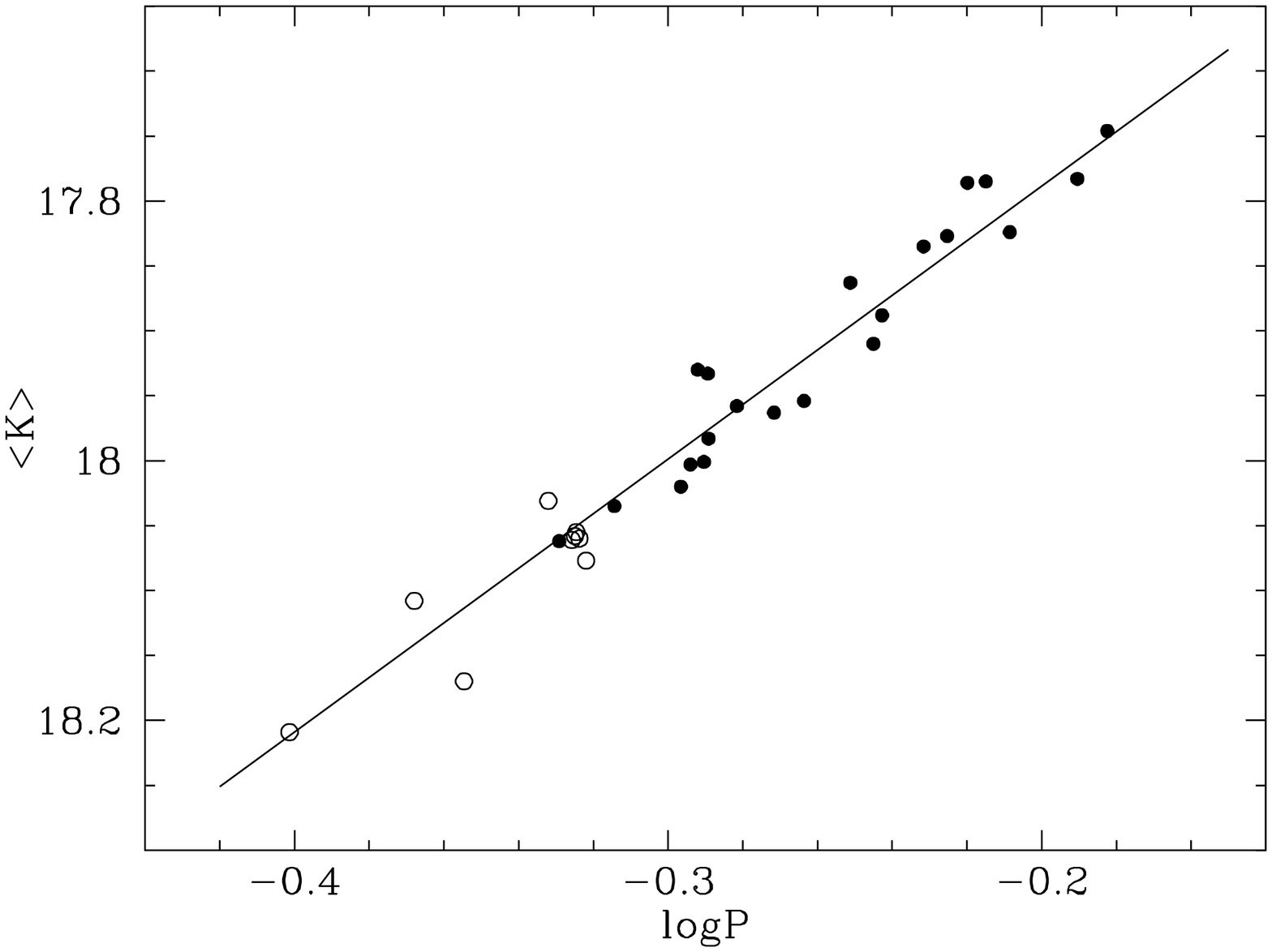}
\caption{The logP-$K$ relation for the Reticulum RR Lyrae stars. 
Open symbols show the \RRc stars after their periods have been 
fundamentalized by adding 0.127 to $\log P$. Filled symbols are
\RRab stars and the straight line represents the theoretical prediction
from Bono et al. (2003) for the derived distance modulus.\label{fig1}}
\end{figure}






\clearpage
\begin{deluxetable}{rccccc}
\tabletypesize{\scriptsize} 
\tablecaption{The weighted intensity averaged $K$-magnitudes based 
on the template fit method of Jones et al. (1996) and the $K$-band 
photometry presented here. The epochs are in Heliocentric Julian Days. 
Type $ab$ and $c$ are for fundamental and first overtone RR Lyrae, 
while type $d$ for candidate mixed-mode variables.\label{tab1}}
\tablewidth{0pt}
\tablehead{
\colhead{ID}  & 
\colhead{$\log P$} & 
\colhead{Epoch} &
\colhead{$\langle K \rangle$} & 
\colhead{$\sigma_K$} & 
\colhead{Type}  \\ 
\colhead{} & 
\colhead{days} & 
\colhead{HJD} & 
\colhead{mag} & 
\colhead{mag} & 
\colhead{} 
}
\startdata
   4 & $-0.45195$ & 2448206.006 & 18.058 & 0.017 &   c \\
   7 & $-0.20853$ & 2448206.070 & 17.824 & 0.015 &  ab \\
  25 & $-0.29657$ & 2448206.125 & 18.020 & 0.018 &  ab \\
  35 & $-0.27164$ & 2448206.285 & 17.963 & 0.015 &  ab \\
  36 & $-0.48161$ & 2448206.220 & 18.170 & 0.018 &   c \\
  37 & $-0.29401$ & 2448206.010 & 18.003 & 0.016 &  ab \\
  38 & $-0.29043$ & 2448206.968 & 18.001 & 0.020 &  ab \\
  41 & $-0.45157$ & 2448206.150 & 18.055 & 0.017 &   d \\
  49 & $-0.28158$ & 2448206.708 & 17.958 & 0.017 &  ab \\
  57 & $-0.28938$ & 2448206.055 & 17.933 & 0.017 &  ab \\
  64 & $-0.28918$ & 2448206.250 & 17.983 & 0.018 &  ab \\
  67 & $-0.18245$ & 2448206.515 & 17.746 & 0.014 &  ab \\
  72 & $-0.45073$ & 2448206.070 & 18.060 & 0.018 &   d \\
  77 & $-0.45276$ & 2448206.258 & 18.061 & 0.019 &   c \\
  80 & $-0.24272$ & 2448206.295 & 17.888 & 0.015 &  ab \\
  97 & $-0.22529$ & 2448206.245 & 17.827 & 0.016 &  ab \\
  98 & $-0.44894$ & 2448206.000 & 18.077 & 0.018 &   d \\
  99 & $-0.21497$ & 2448206.535 & 17.785 & 0.014 &  ab \\
 100 & $-0.23162$ & 2448206.360 & 17.835 & 0.016 &  ab \\
 108 & $-0.32917$ & 2448206.100 & 18.062 & 0.016 &  ab \\
 110 & $-0.45906$ & 2448206.100 & 18.031 & 0.017 &   d \\
 112 & $-0.25124$ & 2448206.250 & 17.863 & 0.019 &  ab \\
 117 & $-0.29205$ & 2448206.245 & 17.930 & 0.017 &  ab \\
 135 & $-0.19047$ & 2448206.510 & 17.783 & 0.014 &  ab \\
 137 & $-0.26363$ & 2448206.455 & 17.954 & 0.016 &  ab \\
 142 & $-0.24504$ & 2448206.200 & 17.910 & 0.018 &  ab \\
 145 & $-0.21988$ & 2448206.338 & 17.786 & 0.018 &  ab \\
 146 & $-0.31439$ & 2448206.044 & 18.035 & 0.019 &  ab \\
 151 & $-0.49493$ & 2448206.134 & 18.108 & 0.019 &   c \\
 181 & $-0.52832$ & 2448206.122 & 18.209 & 0.018 &   c \\
\enddata
\end{deluxetable}

\clearpage
\begin{deluxetable}{lccccccccc}
\tabletypesize{\scriptsize} 
\tablecaption{Comparison of distance moduli based on RR Lyrae stars for 
different GGCs.\label{tab2}}
\tablewidth{0pt}
\tablehead{
\colhead{ID}  & 
\colhead{$[Fe/H]^a$} & 
\colhead{$E(B-V)^b$} &
\colhead{$\langle V \rangle^c$} & 
\colhead{$\sigma_V$} & 
\colhead{$\mu_0^V(BW)^d$}  &
\colhead{$\mu_0^V(FOBE)^e$}  &
\colhead{$(\mu_0^K)^f$} & 
\colhead{$\Delta\mu_0^g$} &  
\colhead{$\Delta\mu_0^h$}  
}
\startdata
NGC~5139~~$\omega$ Cen&-1.62&0.12& 14.57&0.12&$13.57\pm0.16$&\ldots\ldots  &$13.77\pm0.04$& 0.20 & \ldots\\
NGC~5272~~M~3         &-1.57&0.01& 15.61&0.12&$14.94\pm0.16$&$14.97\pm0.07$&$15.15\pm0.06$& 0.21 & 0.18\\
NGC~5904~~M~5         &-1.27&0.03& 15.06&0.08&$14.26\pm0.13$&$14.28\pm0.07$&$14.37\pm0.09$& 0.11 & 0.09\\
NGC~7078~~M~15        &-2.26&0.10& 15.82&0.08&$15.01\pm0.13$&$15.17\pm0.07$&$15.32\pm0.10$& 0.30 & 0.15\\
\enddata
\tablenotetext{a}{Mean cluster metallicity according to Harris (1996).\\
\hspace*{2.5mm} $^b$Cluster reddening according to Harris (1996).\\
\hspace*{2.5mm} $^c$Mean visual magnitude according to: $\omega$ Cen, Olech 
et al. (2003); M~3, Corwin \& Carney (2001); M~5, Caputo et al. (1999); 
M~15, Silbermann, \& Smith (1995).\\
\hspace*{2.5mm} $^d$Distance moduli based on the Baade-Wesselink calibration 
provided by Fernley (1994).\\
\hspace*{2.5mm} $^e$Distance moduli based on First Overtone Blue Edge method 
suggested by Caputo et al. (2000).\\  
\hspace*{2.5mm} $^f$Distance moduli based on K-band mean magnitudes for RR 
Lyrae stars collected by L90 and the semi-empirical relation derived by B03.\\ 
\hspace*{2.5mm} $^g$Difference in distance moduli between the $PLZ_K$ relation 
and the BW method.\\ 
\hspace*{2.5mm} $^h$Difference in distance moduli between the $PLZ_K$ relation 
and the FOBE method.}

\end{deluxetable}


\end{document}